\documentclass[12pt,preprint]{aastex}

\begin{document}
\title{Optical and $\gamma$-ray emissions from internal forward-reverse shocks: application to GRB 080319B?}
\author{Y. W. Yu$^{1,2}$,  X. Y. Wang$^1$, and Z. G. Dai$^1$}
\affil{$^1$Department of Astronomy, Nanjing University, Nanjing
210093, China; \\yuyw, xywang, dzg@nju.edu.cn
\\$^2$Institute of Astrophysics, Huazhong Normal University, Wuhan
430079, China }

\begin{abstract}
In the popular internal shock model for the prompt emission of
gamma-ray bursts (GRBs), collisions between a series of relativistic
shells generate lots of paired forward and reverse shocks. We show
that the synchrotron emission produced by the forward and reverse
shocks respectively could peak at two quite different energy bands
if the Lorentz factors of these two types of shocks are
significantly different with each other (e.g., one shock is
relativistic and the other is Newtonian). We then investigate
whether this scenario is applicable to the case of GRB 080319B and
find that a bimodal distribution of the shell Lorentz factors,
peaking at $\sim400$ and $\sim10^5$, is required. In addition, this
scenario predicts an accompanying inverse-Compton (IC) GeV emission
with a luminosity comparable to (not much higher than) that of the
synchrotron MeV emission, which can be tested with future
\textit{Fermi} observations.
\end{abstract}
\keywords{gamma rays: bursts --- radiation mechanisms: nonthermal}

\section{Introduction}
Since the pioneering works by Rees \& M\'esz\'aros (1994) and
Paczy\'nski \& Xu (1994), it has been widely argued that the prompt
emission of GRBs arises from internal shocks in a relativistic
fireball that consists of a series of shells with different Lorentz
factors and that the observed $\gamma$-ray emission is usually
attributed to synchrotron or inverse-Compton (IC) emission from
power-law electrons in the shocks (M\'esz\'aros \& Rees 1997).
Especially, in the case of IC $\gamma$-ray emission, we can also
expect a bright synchrotron low-energy (e.g., optical) emission as
seed photons for IC scattering. Based on this synchrotron and
synchrotron self-Compton (SSC) emission scenario, M\'esz\'aros \&
Rees (1999) accounted for the bright optical flash of GRB 990123.

In this paper we put forward an alternative model, in which prompt
optical and $\gamma$-ray emissions may be produced only by
synchrotron radiation in internal shocks. It is natural to expect
that internal shocks should consist of paired forward and reverse
shocks, which are produced simultaneously by collisions between the
shells in the fireball. In the case where the Lorentz factors of
these two types of shocks are quite different, two groups of
electrons accelerated by a forward and a reverse shock respectively
during one collision are expected to reach different characteristic
energies. Therefore, by combining the two synchrotron components
contributed by these two shocks, we can get a bimodal photon
spectrum, which might be able to account for some GRBs that have two
spectral components. In particular, under some peculiar conditions,
an optical peak can be found to fit the prompt optical emission of
some GRBs such as the naked-eye GRB 080319B.

For GRB 080319B, with its prompt $\gamma$-ray fluence
$(6.13\pm0.13)\times10^{-4}~\rm erg~cm^{-2}$ (20 keV $-$ 7 MeV;
Racusin et al. 2008b) and redshift $z=0.937$ (Vreeswijk et al.
2008), its isotropic equivalent $\gamma$-ray energy release is
estimated to be $E_{\gamma,iso}=1.4\times10^{54}\rm ergs$
($20\,{\rm keV}-7\,{\rm MeV}$), which is among the highest ever
measured. More surprisingly, this GRB was found to be associated
by an extraordinarily bright optical flash peaked at a visual
magnitude of 5.3 (Racusin et al. 2008b), which is visible even for
the unaided eye. Compared with the extrapolation of the
$\gamma$-ray spectrum to the optical band, the observed flux
density of the optical flash ($\sim20$ Jy) is about ten thousands
times higher. In addition, thanks to the high time resolution, a
fluctuating structure can be seen clearly in the optical light
curve.

For this rare multi-wavelength prompt emission, any scenario
invoking a single synchrotron component from only one emitting
region is obviously not viable. Moreover, different from some other
bright optical flashes such as those of GRBs 990123 and 061007, an
external shock origin for the optical flash of GRB 080319B is
disfavored due to the short duration of the optical pulse and the
lack of an increasing optical pulse duration throughout the whole
prompt phase, as argued by Kumar \& Panaitescu (2008). Instead, an
internal dissipation origin may be plausible. As argued by
M\'esz\'aros \& Rees (1999) for GRB 990123, Kumar \& Panaitescu
(2008) and Racusin et al. (2008b) explained the prompt emission of
GRB 080319B in the synchrotron and SSC emission scenario. However,
as a direct consequence, they predicted a remarkably strong GeV
emission, whose luminosity is about 10 times higher than the
observed MeV one. In contrast, for several bright GRBs detected by
EGRET on {\em CGRO}, the GeV fluence is not higher than that in the
MeV BATSE energy band (e.g. Sommer et al. 1994; Hurley et al. 1994).

However, in our internal shock-produced two-component synchrotron
emission scenario, the problem of GeV emission excess can be
avoided, whereas the prompt optical and MeV $\gamma$-ray fluxes of
GRB 080319B can be still explained if an unusually high variability
of Lorentz factors exists in the fireball. Moreover, since internal
shocks can take place many times, we can naturally understand the
fluctuating structure of the optical and $\gamma$-ray light curves.
However, the emissions in the two energy bands are not strictly
correlated with each other, since the durations of the shocks
determined by the widths of the shells could be different and the
specific shape of the light curves is dependent on the specific
structure of the shells. For simplicity, these complications will be
not considered in our simple model.

Compared with the internal shock model suggested by M\'esz\'aros \&
Rees (1999) for GRB 990123, our work differs from theirs in two
aspects. First, as discussed above, we suggest that both optical and
$\gamma$-ray emissions arise from synchrotron emission, differing
from the synchrotron plus SSC scenario proposed by them. Second, in
their works, only nonrelativistic to semirelativistic internal
shocks are considered, which are produced by collisions between
shells with similar Lorentz factors, while in our paper some
relativistic internal (reverse) shocks would be involved in order to
fit the observations of GRB 080319B.

This paper is organized as follows: in \S2, we describe the dynamics
and synchrotron emission of internal forward-reverse shocks and the
model parameters are expressed as functions of some observational
quantities. In \S3, we constrain the model parameters by the
observations of GRB 080319B and then some implications from these
results are discussed. In addition, the contribution to the prompt
emission by IC scattering of the electrons is also considered with
the inferred model parameters. Finally, conclusions and discussion
are given in \S4.

\section{Two-component synchrotron emission from internal shocks}
\subsection{The dynamics and electron energy distributions}
In the internal shock model, the central engine of GRBs is assumed
to eject a fireball consisting of a series of shells with different
Lorentz factors $\gamma_{\rm shell}$ during the prompt phase.
Considering two shells that are ejected subsequently, for example,
if the posterior shell (denoted by 4) moves more rapidly than the
prior one (denoted by 1), a collision takes place at radius
$R_{is}\approx2\gamma_{\rm 1}^2c\delta t\phi_z^{-1}$ (Yu \& Dai
2008) to produce an emission pulse, where $\delta t$ is the observed
variability time and $\phi_z=1+z$ is introduced due to the
cosmological dilution of time. Because of the collision, a pair of
shocks (i.e., internal shocks) could arise: a forward shock
propagating into shell 1 and a reverse shock propagating into shell
4. The shocked regions in shells 1 and 4 are denoted by 2 and 3,
respectively.

According to the jump conditions between the two sides of a shock
(Blandford \& McKee 1976), we can calculate the comoving internal
energy densities of the two shocked regions by
$e_2=(\gamma_{21}-1)(4\gamma_{21}+3)n_1m_pc^2$ and $
e_3=(\gamma_{34}-1)(4\gamma_{34}+3)n_4m_pc^2$, where $\gamma_{21}$
or $\gamma_{34}$ is the Lorentz factor of region 2 or 3 relative to
the unshocked region 1 or 4. The comoving proton number density
$n_i$ of unshocked region $i$ can be calculated by $n_{i}={L_{k,i}/(
4\pi R_{is}^2\gamma_{i}^2m_pc^3})$ for an isotropic kinetic-energy
luminosity $L_{k,i}$ and a bulk Lorentz factor $\gamma_i~(\gg1)$ of
the unshocked shell. The mechanical equilibrium between the two
shocked regions requires $e_2=e_3$, which yields
\begin{equation}
{(\gamma_{21}-1)(4\gamma_{21}+3)\over(\gamma_{34}-1)(4\gamma_{34}+3)}={n_4\over
n_1}=\left({\gamma_1\over \gamma_4}\right)^2,\label{dyne}
\end{equation}
where $L_{k,1}=L_{k,4}=L_k$ is supposed. By assuming
$\gamma_4\gg\gamma_1$, the above equation leads to (Yu \& Dai 2008)
\begin{equation}
\gamma_{21}-1={1\over2}\left({\gamma_1\over\gamma}+{\gamma\over\gamma_1}\right)-1\approx\xi\ll1,
~~\gamma_{34}={1\over2}\left({\gamma\over\gamma_4}+{\gamma_4\over\gamma}\right)\approx{\gamma_4\over2\gamma},
\end{equation}
where $\gamma=\gamma_1(1+\sqrt{2\xi})$ is the Lorentz factor of the
shocked regions. This indicates that the reverse shock is
relativistic and the forward shock is Newtonian. Moreover, following
Dai \& Lu (2002), the total number of the electrons swept-up by the
forward and reverse shocks during a period of $\delta t$ can be
expressed by $N_2={2\sqrt{2\xi} L_k\delta
t/\left(\phi_z\gamma_1m_pc^2\right)}$ and $N_3={L_k\delta
t/\left(\phi_z\gamma_4m_pc^2\right)}$, respectively\footnote{In
these expressions, the possible spreading of the shells (i.e.
decreasing of $L_k$) is not taken into account, since a detailed
description for the dynamic evolution is not necessary for our
calculations. In this paper, the constant $L_k$ can be seen as an
effective kinetic-energy luminosity.}.

Both forward and reverse shocks can accelerate particles to high
energies and amplify magnetic fields. As usual, we assume that the
energies of the hot electrons and magnetic fields are fractions
$\epsilon_e$ and $\epsilon_B$ of the total internal energy,
respectively. Thus, the strength of the magnetic fields is given
\begin{equation}
B_i=\left(8\pi
\epsilon_{B}e_i\right)^{1/2}={1\over\sqrt{2}c^{3/2}}\left({\phi_z^2\epsilon_{B}L_k\over\delta
t^2\gamma^{6}}\right)^{1/2},~~i=2,3.
\end{equation}
For the shock-accelerated electrons, a power-law energy
distribution, $dn_e/d\gamma_e\propto\gamma_e^{-p}$ for
$\gamma_e\geq\gamma_{e,m}$, is assumed. The characteristic random
Lorentz factors of these hot electrons in regions 2 and 3 are
determined respectively by
\begin{eqnarray}
\gamma_{e,m,2}=\hat{\epsilon}_{e}{m_p\over m_e}(\gamma_{21}-1)\approx{\hat{\epsilon}_{e}}{m_p\over
m_e}\xi,\\
\gamma_{e,m,3}=\hat{\epsilon}_{e}{m_p\over
m_e}(\gamma_{34}-1)\approx{\hat{\epsilon}_{e}\over2}{m_p\over
m_e}{\gamma_{4}\over\gamma},\label{gem}
\end{eqnarray}
where $\hat{\epsilon}_{e}=\epsilon_{e}(p-2)/(p-1)$. That
$\gamma_{e,m,3}\gg\gamma_{e,m,2}$, due to $\gamma_4/\gamma\gg2\xi$,
indicates that the characteristic energy of the reverse-shocked
electrons is much higher than that of the forward-shocked electrons.
Therefore, the resulting synchrotron photons emitted by these two
types of electrons are expected to peak at two different energy
bands and thus two distinct spectral components would be observed as
in GRB 080319B. To be specific, the reverse shock is responsible for
emission in a relatively high-energy band, while the forward shock
contributes to a relatively low-energy component.

In both shocked regions, the hot electrons with energies above
$\gamma_{e,c,i}m_ec^2$ lose most of their energies during a cooling
time $t_{c,i}$, where the cooling Lorentz factor is determined by
${\gamma}_{e,c,i}={6\pi m_ec\phi_z/\left( y_i\sigma_T{B}_i^2\gamma
t_{c,i}\right)}$. The parameter $y_i$, defined as the ratio of the
total luminosity to the synchrotron one, is introduced by
considering the cooling effect due to the IC emission besides the
synchrotron cooling. As pointed out by Ghisellini et al. (2000), the
theoretical synchrotron spectrum arising from these electrons,
calculated by using the standard assumption that the magnetic field
maintains a steady value throughout the shocked region, leads to a
spectral slope $F_\nu \propto \nu^{-1/2}$ below $\sim$100 keV, which
is in contradiction to the much harder spectra observed. In order to
overcome this problem, Pe'er \& Zhang (2006) suggested that the
magnetic field created by a shock could decay on a length scale
($\lambda_{B,i}$) much shorter than the comoving width ($\Delta_i$)
of the shocked region, i.e., $\lambda_{B,i}=\Delta_i/f_{B,i}$
($f_{B,i}\gg1$). In other words, the shocked region can be roughly
divided into a magnetized part immediately after the shock front and
a further unmagnetized part. Under this assumption, the cooling time
of the electrons should be determined by the time during which the
electrons traverse the magnetized region, i.e., $t_{c,i}=\delta
t/f_{B,i}$. Although the size of the magnetized region is reduced
significantly by the field-decay effect (as found in \S3.3), this
region could be still wide enough for electrons to lose a great part
of their energy when they traverse it. In this case, the cooling
Lorentz factor $\gamma_{e,c,i}$ of electrons is not much higher than
$\gamma_{e,m,i}$, so that the radiation efficiency of the electrons
is not reduced drastically compared to the case without any magnetic
field decay.

\subsection{Two-component synchrotron emission}
With the electron distributions and the magnetic fields described
above, we can give the resultant synchrotron spectra using the
method developed by Sari et al. (1998). The reference peak energies
of the synchrotron spectra of the forward and reverse shocks are
taken to be in optical and $\gamma$-ray bands, respectively. Then,
the model parameters can be expressed as functions of some
observational quantities.

For the electrons in the magnetized reverse-shocked region, two
break frequencies of the  synchrotron spectrum are given by
\begin{eqnarray}
\nu_{m,3} & = & { q_e\over2\pi
m_ec\phi_z}{\gamma}_{e,m,3}^2{B}_3\gamma \nonumber \\
& = & {\sqrt{2}m_p^2q_e\over16\pi
m_e^3c^{5/2}}\left({\epsilon_{B}\hat{\epsilon}_e^4L_k\gamma_4^4}\over\delta
t^2\gamma^8\right)^{1/2},\label{num,3}
\end{eqnarray}
\begin{eqnarray}
\nu_{c,3} & = & {q_e\over2\pi
m_ec\phi_z}{\gamma}_{e,c,3}^2{B}_3\gamma \nonumber \\
& = & {36\sqrt{2}\pi m_eq_ec^{11/2}\over \sigma_{\rm T}^2}\left({\delta t^2f_{B,3}^4\gamma^{16}}\over
\phi_z^4y_3^4\epsilon_{B}^3L_k^3\right)^{1/2}.\label{nuc,3}
\end{eqnarray}
The peak flux density of the spectrum at $\nu_{\rm
p}=\min[\nu_m,\nu_c]$ reads
\begin{eqnarray}
F_{\nu,{\rm p},3} & = & {\phi_z\over4\pi d_{L}^2}{
m_{e}c^2\sigma_{T}\over 3q_e}{N_{3}\over
f_{B,3}}B_{3}\gamma\nonumber \\& = & {\sqrt{2}m_e\sigma_{\rm
T}\over24\pi m_pq_ec^{3/2}}\left({\phi_z^2\epsilon_{B}L_k^3}\over
d_L^4f_{B,3}^2\gamma^4\gamma_4^2\right)^{1/2},\label{fmax,3}
\end{eqnarray}
where the parameter $f_{B,3}$ is introduced because, at any moment,
only a fraction $1/f_{B,3}$ of the total reverse-shocked electrons
locate at the magnetized region and other electrons in the
unmagnetized region do not contribute to the synchrotron emission.
The quantities in the left sides of equations
(\ref{num,3})-(\ref{fmax,3}) can be inferred from an observed prompt
$\gamma$-ray spectrum, while the right sides are functions of the
model parameters. We can therefore solve these equations to find the
values of some model parameters,
\begin{eqnarray}
L_k&=&2.5\times10^{53}{\rm
erg~s^{-1}}~{y_{3,0}\hat{\epsilon}_{e,-1}^{-1}}\nonumber\\
&&\times\left(d_{L,28}^{2}{F_{\nu,{\rm
p},3,-25}}^{}\nu_{c,3,20}^{1/2}\nu_{m,3,20}^{1/2}\right)\nonumber\\
& \equiv & {{y_3\over\hat{\epsilon}_{e}}}L_k^*\label{lk},
\\
\gamma_{4}&=&4\times10^4~\gamma_{2.5}^{2}{y_{3,0}^{-1/4}\epsilon_{B,-1}^{-1/4}\hat{\epsilon}_{e,-1}^{-3/4}}\nonumber\\
&&\times\left( {\delta t_{0}}^{1/2}d_{L,28}^{-1/2}F_{\nu,{\rm
p},3,-25}^{-1/4}\nu_{c,3,20}^{-1/8}\nu_{m,3,20}^{3/8}\right),
\\
f_{B,3}&=&6\times10^{3}~
\gamma_{2.5}^{-4}{y_{3,0}^{7/4}\epsilon_{B,-1}^{3/4}\hat{\epsilon}_{e,-1}^{-3/4}}\nonumber\\
&&\times\left({\delta
t_0}^{-1/2}\phi_{z,0.3}d_{L,28}^{3/2}F_{\nu,{\rm
p},3,-25}^{3/4}\nu_{c,3,20}^{7/8}\nu_{m,3,20}^{3/8}\right),
\end{eqnarray}
where and hereafter the convention $Q=10^xQ_x$ is adopted in cgs
units. The quantities in the brackets are basically determined by
the observational data and the values of $L_k$, $\gamma_4$, and
$f_{B,3}$ are modulated by the remaining free parameters. In
particular, $\gamma_4$ and $f_{B,3}$ are strongly dependent on
$\gamma$ that can be constrained by optical observations. The fact
that $L_{k}$ is independent of the parameter $f_{B,3}$ indicates
that the hypothesis of magnetic field decay does not increase the
energy requirement of the model. This is because all the
reverse-shocked electrons, when they traverse the tiny magnetized
region at different times, have released a great part of their
energy to $\gamma$-rays (as indicated by $\nu_{m,3}\sim\nu_{c,3}$)
via synchrotron emission. The observed $\gamma$-ray emission is
mainly contributed by the emission from this tiny region behind the
shock front, no matter whether the other part of the shocked region
is magnetized or not.

In order to study the properties of the forward shock, we now focus
on a possible low-energy emission, whose spectral information is
however not as rich as the $\gamma$-ray component. We calculate the
peak frequency of the synchrotron spectrum produced by the forward
shock as
\begin{eqnarray}
\nu_{m,2} & = & {q_e\over2\pi
m_ec\phi_z}{\gamma}_{e,m,2}^2{B}_2\gamma \nonumber
\\ & = & {\sqrt{14}m_p^2q_e\over4\pi
m_e^3c^{5/2}}\left({\epsilon_{B}\hat{\epsilon}_e^4L_k\xi^{5}}\over \delta t^2\gamma^4\right)^{1/2}.\label{num,2}
\end{eqnarray}
In order to explain the optical spectrum measured by Raptor after 80
s for GRB 080319B (Kumar \& Panaitascu 2008; Wo\'zniak et al. 2008)
in the next section, we assume that this peak frequency is below the
optical band (i.e., $\nu_{m,2}<\nu_o\equiv5\times10^{14}\rm ~Hz$),
which yields
$\gamma>\tilde{\gamma}\equiv288~y_3^{1/4}\epsilon_{B,-1}^{1/4}\hat{\epsilon}_{e,-1}^{3/4}\delta
t_0^{-1/2}L_{52.4}^{*1/4}$. This requirement is however not
necessary for common GRBs. For $\nu_{m,2}<\nu_o<\nu_{c,2}$, we can
calculate the optical flux density and the synchrotron
self-absorption thickness at $\nu_o$ by
\begin{eqnarray}
F_{\nu,o}=F_{\nu,\rm
p,2}\left({\nu_o\over\nu_{m,2}}\right)^{-(p-1)/2},\label{fo2}
\end{eqnarray}
\begin{eqnarray}
\tau_{sa,o}=\tau_{sa,m}\left({\nu_o\over\nu_{m,2}}\right)^{-(p+4)/2},\label{tao2}
\end{eqnarray}
where
\begin{eqnarray}
F_{\nu,{\rm p},2} & = & {\phi_z\over4\pi d_{L}^2}{ m_{e}c^2\sigma_{T}\over 3q_e}{N_{2}\over f_{B,2}}B_{2}\gamma
\nonumber \\&=& {\sqrt{7}m_e\sigma_{\rm T}\over6\pi m_pq_ec^{3/2}}\left({\phi_z^2\epsilon_{B}L_k^3\xi^2}\over
d_L^4f_{B,2}^2\gamma^6\right)^{1/2},
\end{eqnarray}
\begin{eqnarray}
\tau_{sa,m} & = & {5q_e\over B_2\gamma_{e,m,2}^5}{N_2\over 4\pi
R^2f_{B,2}}\nonumber
\\ & = & {5\sqrt{7}m_e^5q_e\over28\pi
m_p^6c^{5/2}}\left({}L_k\over\epsilon_{B}\hat{\epsilon}_e^{10}f_{B,2}^2\gamma^4\xi^{10}\right)^{1/2}.
\end{eqnarray}
Combining equations (\ref{num,2}), (\ref{fo2}) and (\ref{tao2}), we
write the parameters $\gamma$ and $f_{B,2}$ as functions of
$\epsilon_B$ and $\hat{\epsilon}_e$,
\begin{eqnarray}
\gamma&=&420~y_{3,0}^{1/16}\epsilon_{B,-1}^{1/16}\hat{\epsilon}_{e,-1}^{-1/16}\nonumber\\
&&\times\left(\delta
t_0^{-5/8}\phi_{z,0.3}^{-1/4}d_{L,28}^{1/2}F_{\nu,o,-22}^{1/4}L_{k,52.4}^{*1/16}\tau_{sa,o,-1}^{-1/4}\right),\\
f_{B,2}&=&180~y_{3,0}^{3(6+p)/16}\epsilon_{B,-1}^{(2+3p)/16}\hat{\epsilon}_{e,-1}^{-(34-13p)/16}\nonumber
\\& & \times\left[\delta
t_0^{(14+p)/8}\phi_{z,0.3}^{(6+p)/4}d_{L,28}^{-(6+p)/2}\right.\nonumber\\
&&\times\left.F_{\nu,o,-22}^{-(6+p)/4}L_{k,52.4}^{*{3(6+p)/16}}\tau_{sa,o,-1}^{(2+p)/4}\right],
\end{eqnarray}
where the kinetic-energy luminosity $L_k=2.5\times10^{52}{\rm
erg~s^{-1}}(y_3/\hat{\epsilon}_{e})L^*_{k,52.4}$ obtained in
equation (\ref{lk}) has been substituted. The value of $\gamma$ is
mainly determined by the observational data and insensitive to the
remaining free parameters $\epsilon_B$ and $\hat{\epsilon}_e$. For
common GRBs, the value of $\tau_{sa,o}$ is thought to be very high
(e.g., $\sim10^3$) and thus the value of $\gamma$ could be lower
than 100, which leads to hundreds for $\gamma_4$. However, for GRB
080319B, $\tau_{sa,o}$ is deemed to be not larger than unity to
ensure a bright optical flash and we suggest $\tau_{sa,o}=0.1$ as a
reference value hereafter.

\section{Application to GRB 080319B}
GRB 080319B triggered the Swift-Burst Alert Telescope (15-350 keV)
at $T_0$=06:12:49 UT on March 19, 2008 (Racusin et al. 2008a) and
was simultaneously detected by the Konus $\gamma$-ray detector
onboard the Wind satellite (20 keV-15 MeV; Golenetskii et al. 2008).
The time-averaged Konus-Wind $\gamma$-ray spectrum can be fitted
well by the Band function (Band et al. 1993) with a low energy slope
of $0.855^{+0.014}_{-0.013}$ below the peak of $E_p=675\pm22$ keV
and a high-energy slope of $-3.59^{+0.32}_{-0.62}$ above the peak
(Racusin et al. 2008b). The burst had a peak flux of
$F_p=(2.26\pm0.21)\times10^{-5}\rm erg~cm^{-2}~s^{-1}$ and thus the
peak isotropic equivalent luminosity was
$L_{p,iso}=(1.01\pm0.09)\times10^{53}\rm erg~s^{-1}$. Using the
values of $F_p$ and $E_p$, we roughly estimate the peak flux density
of the $\gamma$-ray spectrum, $F_p/E_p\approx14$ mJy. Compared with
the extrapolation of the $\gamma$-ray spectrum to the optical band,
the observed flux density of the optical flash ($\sim20$ Jy) is
about ten thousands times higher.

\subsection{The model parameters}
Adopting $z=0.937$ ($d_{L}=1.9\times10^{28}\rm cm$), $\delta t\sim
3$ s, $F_{\nu,o}\sim 20 \rm Jy$ (at $\nu_o=5\times10^{14}\rm Hz$),
$F_{\nu,\rm p,3}\sim14\rm mJy$, $h\nu_{m,3}\sim 675\rm keV$ and
denoting $\nu_{c,3}\equiv x\times \nu_{m,3}\equiv 10^0x_0\times
\nu_{m,3} $ for GRB 080319B, we derive the model parameters,
\begin{eqnarray}
L_k&\simeq &2\times10^{54}~{\rm
erg~s^{-1}}~x_0^{1/2}y_{3,0}\hat{\epsilon}_{e,-1}^{-1},\label{para1}\\
\gamma&\simeq&400~x_0^{1/32}y_{3,0}^{1/16}\epsilon_{B,-1}^{1/16}\hat{\epsilon}_{e,-1}^{-1/16},\label{para2}\\
\gamma_4&\simeq&9\times10^4~x_0^{-1/16}y_{3,0}^{-1/8}\epsilon_{B,-1}^{-1/8}\hat{\epsilon}_{e,-1}^{-7/8},\label{para3}\\
f_{B,2}&\simeq&700~x_0^{3(6+p)/32}y_{3,0}^{3(6+p)/16}\epsilon_{B,-1}^{(2+3p)/16}\hat{\epsilon}_{e,-1}^{-(34-13p)/16},\label{para4}\\
f_{B,3}&\simeq&7\times10^3~x_0^{3/4}y_{3,0}^{3/2}\epsilon_{B,-1}^{1/2}\hat{\epsilon}_{e,-1}^{-1/2},\label{para5}
\end{eqnarray}
where a fiducial value of unity is assumed for $y_3$, which will be
proved in \S 3.2. Besides a constraint by the maximum allowed
equipartition value ($\epsilon_e\lesssim 0.3$ and
$\epsilon_B\lesssim 0.3$), the remaining free parameters
$\epsilon_B$ and $\hat{\epsilon}_e$ satisfy
$\hat{\epsilon}_{e}<0.09(xy_3^2\epsilon_{B}^2)^{-3/26}$ given
$\gamma>\tilde{\gamma}$. The upper limit of $\hat{\epsilon}_e$ is
insensitive to the value of $\epsilon_B$ (strictly, with a decrease
of $\epsilon_B$, the upper limit of $\hat{\epsilon}_e$ increases
slightly). Taking $\hat{\epsilon}_e<0.09$ ($\epsilon_e\lesssim 0.3$)
as a conservative estimate, we find:

(i) $L_k\gtrsim2\times10^{54}\rm erg~s^{-1}$. This is a natural
result due to the high observed $\gamma$-ray luminosity
($\sim10^{53}\rm erg~s^{-1}$) of GRB 080319B. The MeV $\gamma$-ray
radiation efficiency of the reverse shock can be estimated by
$\eta\sim0.05x_0^{-1/2}y_{3,0}^{-1}\hat{\epsilon}_{e,-1}$. We next
calculate the total isotropic-equivalent energy release of GRB
080319B,
\begin{eqnarray}
E_{k,iso}=2E_{\gamma,iso}/\eta\sim5\times10^{55}x_0^{-1/2}y_{3,0\hat{\epsilon}_{e,-1}^{-1}}~\rm ergs,
\end{eqnarray}
where a factor 2 is introduced by considering a similar amount of
energy carried by the forward shocks. Using a very small jet angle
$\theta_j=0.2^o$ that is found by Racusin et al. (2008b), we get the
beaming-corrected energy release of GRB 080319B,
\begin{eqnarray}
E_{k,jet}=E_{k,iso}\theta_j^2/2\sim10^{51}x_0^{-1/2}y_{3,0}\hat{\epsilon}_{e,-1}^{-1}~\rm ergs,
\end{eqnarray}
which is a typical value for common GRBs.

(ii) $\gamma\sim400$. This is a typical value for the Lorentz factor
of merged GRB ejecta after internal shocks. Thus, we can obtain the
internal shock radius for GRB 080319B,
\begin{eqnarray}
R_{is}\approx2\gamma_{\rm 1}^2c\delta t\phi_z^{-1}\approx10^{16}{\rm
cm}~\delta t_{0.5}\phi_{z,0.3}^{-1}\gamma_{1,2.6}^{2} ,
\end{eqnarray}
which is larger than those of common GRBs ($\sim10^{14}$cm). So, the
dynamic influence of the circum-stellar medium on the leading shell
($\gamma_1\sim \gamma\sim400$) needs to be assessed. We calculate
the deceleration radius of the leading shell via
\begin{eqnarray}
R_{dec}={L_k\delta t\phi_z^{-1}\over2\pi \gamma_1^2A c^2}\approx5\times10^{16}{\rm cm}~A_{*,-1}^{-1}\delta
t_{0.5}\phi_{z,0.3}^{-1}\gamma_{1,2.6}^{-2}L_{k,54.3},
\end{eqnarray}
where a wind-like circumburst medium ($\rho=Ar^{-2}$ with
$A=5\times10^{11}{\rm g~cm^{-1}}A_*$) is assumed. Moreover, as
claimed by Racusin et al. (2008b), a tenuous wind with an upper
limit of $A_*\sim 0.03$ is required by the afterglow data of GRB
080319B. In this case, the deceleration radius is larger. Therefore,
we conclude that the deceleration of the shells can be ignored at
the times where internal shocks among the shells occur.

(iii) $\gamma_4\gtrsim 10^5$. This high Lorentz factor is allowable
for acceleration of an initial fireball with very low baryon
contamination (Piran 1999).

(iv) $f_{B,3}\gtrsim 10f_{B,2}$. Our constraint on $f_{B,3}$, which
is much larger than unity and much less than $\Delta/\lambda_s$
($\lambda_s$ is the plasma skin depth), is consistent with that
found by Pe'er \& Zhang (2006) for other GRBs. However, the physical
underpinning of these values of $f_{B}$ is unknown. So the
difference in $f_{B}$ for different shocked regions lacks a
reasonable physical explanation and it might be related to different
shock strengths of the forward and reverse shocks.

From the above results, we can see that most of the inferred model
parameters are reasonable and acceptable even in common GRBs, except
for an unusually high variability of Lorentz factors denoted by
$\gamma_4/\gamma_1\sim300$. Although the $\gamma$-ray emission of
GRB 080319B seems to be not unusual, the relatively high value of
$\delta t\sim3$s (versus $\sim$10 ms for common GRBs) implies an
unusually large internal shock radius, which ensures the synchrotron
self-absorption frequency below the optical band and reduces the
magnetic field strength. Therefore, in order to produce sufficiently
energetic $\gamma$-ray photons, it is necessary to invoke some
highly relativistic internal shocks that require high variability of
Lorentz factors in the fireball. Since the GRB central engine is far
from being thoroughly understood, it is difficult to demonstrate
whether the central engine can produce such a drastically varying
fireball or not, but some possible origins can be still imagined. In
the collapsar model, for example, a shell passing through the
envelope of a progenitor star should sweep up and clear away the
envelope material, leaving a channel behind the shell. A following
shell will pass through this clear channel and thus have a very low
baryon contamination. This might lead to a highly relativistic
($\gamma_{\rm shell}\sim10^5$) shell. However, due to lateral
diffusion of the channel wall, this channel will be possibly
contaminated by baryons again some time ($\sim$ order of seconds)
later. Subsequently, such a switching-on-and-off process of the
channel repeats again and again. As a result, relatively slow and
rapid shells are generated alternately. In reality, this process is
unlikely to be so regular because a slow/rapid shell could be
followed by another slow/rapid shell. Therefore, the
temporally-correlated optical and $\gamma$-ray emissions from this
process could be polluted by the emission due to the collisions
between slow-slow or rapid-rapid shells. This along with other
effects (e.g., different shock-crossing times of shells and so on,
see \S4) may lead to the fact that the prompt optical and
$\gamma$-ray emissions are not correlated finely, as observed in GRB
080319B.

To summarize, if some unknown physical processes of the central
engine can give rise to a fireball in which the Lorentz factors and
densities vary drastically in a variability timescale of few
seconds, our internal shock-produced two-component synchrotron
emission scenario may account for some fundamental features of the
optical and MeV $\gamma$-ray emissions of GRB 080319B.

\subsection{Inverse Compton emission}
We denote $\tau_{i}={\sigma_TN_i/\left(4\pi R^2f_{B,i}\right)}$ and
$\tau^{\dag}_{i}={\sigma_TN_i/\left(4\pi R^2\right)}$ as the Thomson
optical depth of the magnetized and unmagnetized regions,
respectively. The SSC emission from the two magnetized regions is
considered first. Following Sari \& Esin (2001) for $p=2.5$, we
estimate the two break energies of the SSC spectrum contributed by
the forward shock by
\begin{eqnarray}
h\nu_{m,2}^{\rm SSC}=2\gamma_{e,m,2}^2h\nu_{m,2}=1.5~{\rm
keV}~x_0^{3/16}y_{3,0}^{3/8}\epsilon_{B,-1}^{3/8}\hat{\epsilon}_{e,-1}^{29/8},
\end{eqnarray}
\begin{eqnarray}
h\nu_{c,2}^{\rm SSC}=2\gamma_{e,c,2}^2h\nu_{c,2}=70~{\rm
GeV}~x_0^{2}y_{2,0}^{-4}y_{3,0}^{4}\hat{\epsilon}_{e,-1}^{2}.
\end{eqnarray}
The peak flux at $\nu_{c,2}^{\rm SSC}$ can be estimated by
\begin{eqnarray}
[\nu F_{\nu}]_{{\rm p},2}^{\rm SSC}&\sim&\nu_{c,2}^{\rm SSC}\tau_2F_{\nu,{\rm p},2}\left({\nu_{c,2}^{\rm
SSC}/\nu_{m,2}^{\rm SSC}}\right)^{-(p-1)/2}\nonumber\\
& \sim &2\times10^{-8} ~{\rm erg~cm^{-2}s^{-1}}
\label{sscflux1}
\end{eqnarray}
For the reverse shock, as the SSC peak enters the Klein-Nishina
regime, the real peak energy can be determined by (Gupta \& Zhang
2007; Fragile et al. 2004)
\begin{eqnarray}
h\nu_{KN,3}^{\rm SSC}={\gamma^2m_e^2c^4\over h\nu_{m,3}}=60~{\rm
GeV}~x_0^{1/16}y_{3,0}^{1/8}\epsilon_{B,-1}^{1/8}\hat{\epsilon}_{e,-1}^{-1/8},
\end{eqnarray}
at which a negligible flux is found from
\begin{eqnarray}
[\nu F_{\nu}]_{{\rm p},3}^{\rm SSC}&\sim&\nu_{KN,3}^{\rm SSC}\tau_3F_{\nu,{\rm p},3}\left({\nu_{KN,3}^{\rm
SSC}\over2\gamma_{e,c,3}^2\nu_{c,3}}\right)^{1/3}\nonumber\\
&\sim &8\times10^{-11} ~{\rm erg~cm^{-2}s^{-1}}. \label{sscflux2}
\end{eqnarray}
According to the definition of parameter $y_i$, we obtain
\begin{eqnarray}
y_2=1+{L^{\rm SSC}_{2}\over L^{\rm syn}_{2}}\lesssim1+{[\nu F_{\nu}]_{\rm p,2}^{\rm SSC}\over[\nu F_{\nu}]_{\rm
opt}}\approx1.2,\\
y_3=1+{L^{\rm SSC}_{3}\over L^{\rm syn}_{3}}\approx1+{[\nu F_{\nu}]_{\rm p,3}^{\rm SSC}\over[\nu F_{\nu}]_{\rm
MeV}}\approx1,
\end{eqnarray}
where $[\nu F_{\nu}]_{\rm opt}=10^{-7}{\rm erg~cm^{-2}s^{-1}}$ and
$[\nu F_{\nu}]_{\rm MeV}=2.3\times10^{-5}~\rm erg~cm^{-2}s^{-1}$.
This indicates that the SSC emission is relatively unimportant and
the estimations in equations (\ref{para1})-(\ref{sscflux2}) under
the assumption, $y_2\sim y_3\sim1$, are self-consistent.

Although a part of the electron energy can be released via
synchrotron and SSC emissions when the electrons traverse the
magnetized regions, about half energy of the reverse-shocked
electrons and almost all energy of the forward-shocked electrons are
still held by the electrons when they enter into the unmagnetized
regions. This remaining energy can be no longer released via
synchrotron radiation because of the lack of magnetic fields, but
can via external inverse Compton (EIC) scattering due to the
existence of the radiation fields. Determined by this EIC cooling,
the cooling Lorentz factor of the electrons in both of the
unmagnetized regions reads
\begin{eqnarray}
{\gamma}^{\dag}_{e,c}={m_ec\phi_z\over {4\over3}\sigma_Tu_{\gamma}\gamma \delta t},
\end{eqnarray}
where the radiation energy density contributed by the synchrotron
radiation from the two magnetized regions can be calculated by
$u_{\gamma}=(y_2+y_3-2)(B^2/8\pi)\equiv Y^{\dag}(B^2/8\pi)$ by
considering $B_2=B_3\equiv B$. Then we have
\begin{eqnarray}
{\gamma}^{\dag}_{e,c}={6\pi m_ec\phi_z\over Y^{\dag}\sigma_TB^2\gamma \delta
t}=15~x_0^{-11/32}{Y_{-0.7}^{\dag-1}}y_{3,0}^{-11/16}\epsilon_{B,-1}^{-11/16}\hat{\epsilon}_{e,-1}^{11/16}.
\end{eqnarray}
and ${\gamma}^{\dag}_{e,c}<(\gamma_{e,m,2},\gamma_{e,m,3})$. For an
EIC spectrum produced by upscattering the seed photons from
magnetized region $j$ by the electrons in unmagnetized region $i$,
we calculate its two characteristic break frequencies by
\begin{eqnarray}
\nu_{L,i}^{(j)}=2{\gamma}_{e,L,i}^2\nu_{L,j},
\end{eqnarray}
\begin{eqnarray}
\nu_{H,i}^{(j)}=2{\gamma}_{e,H,i}^2\nu_{H,j},
\end{eqnarray}
and estimate its peak flux at the peak frequency $\nu_{H,i}^{(j)}$ roughly by
\begin{eqnarray}
\left[\nu F_{\nu}\right]_{{\rm p},i}^{(j)}\sim\nu_{H,i}^{(j)}\tau_i^{\dag}F_{\nu,{\rm
p},(j)}\left({\nu_{H,i}^{(j)}/\nu_{L,i}^{(j)}}\right)^{-(p-1)/2},\label{eicflux}
\end{eqnarray}
where the subscription $L$ represents the low break frequency of the
seed photons and the low break Lorentz factor of the target
electrons, while $H$ represents the high ones. Considering the two
synchrotron components for seed photons and the two population
unmagnetized electrons, four EIC components are expected:

(i) For $i=3$ and $j=2$, we have
\begin{eqnarray}
h\nu_{L,3}^{(2)}=2{{\gamma}^{\dag2}_{e,c}}h\nu_{m,2}=0.5~{\rm
keV}~x_0^{-1/2}Y_{-0.7}^{-2}y_{3,0}^{-1}\epsilon_{B,-1}^{-1}\hat{\epsilon}_{e,-1}^{3},
\end{eqnarray}
\begin{eqnarray}
h\nu_{H,3}^{(2)}=2{{\gamma}^{2}_{e,m,3}}h\nu_{c,2}=6~{\rm
TeV}~x_0^{29/32}y_{2,0}^{-2}y_{3,0}^{29/16}\epsilon_{B,-1}^{-3/16}\hat{\epsilon}_{e,-1}^{19/16}.
\end{eqnarray}
Although the peak frequency $\nu_{H,3}^{(2)}$ is just around the
Klein-Nishina break determined by
$h\nu_{KN,3}^{(2)}={\gamma^2m_e^2c^4/( h\nu_{c,2})}\approx6~{\rm
TeV}$, we may still derive a relatively low flux from equation
(\ref{eicflux}),
\begin{eqnarray}
\left[\nu F_{\nu}\right]_{{\rm p},3}^{(2)}\sim8\times10^{-8}\rm ~erg~cm^{-2}s^{-1}.\label{eicflux3}
\end{eqnarray}

(ii) For $i=3$ and $j=3$, we have
\begin{eqnarray}
h\nu_{L,3}^{(3)}=2{{\gamma}^{\dag2}_{e,c}}h\nu_{c,3}=0.3~{\rm
GeV}~x_0^{5/16}Y_{-0.7}^{-2}y_{3,0}^{-11/8}\epsilon_{B,-1}^{-11/8}\hat{\epsilon}_{e,-1}^{11/8},
\end{eqnarray}
\begin{eqnarray}
h\nu_{H,3}^{(3)}=2{{\gamma}^{2}_{e,m,3}}h\nu_{m,3}=0.6~{\rm
EeV}~x_0^{-3/16}y_{3,0}^{-3/8}\epsilon_{B,-1}^{-3/8}\hat{\epsilon}_{e,-1}^{3/8}.
\end{eqnarray}
The peak frequency $\nu_{H,3}^{(3)}$ is however higher than the
Klein-Nishina break frequency,
$h\nu_{KN,3}^{(3)}={\gamma^2m_e^2c^4/( h\nu_{m,3})}\approx60~{\rm
GeV}$, and thus the real peak flux can be calculated at
$\nu_{KN,3}^{(3)}$ to be
\begin{eqnarray}
\left[\nu F_{\nu}\right]_{{\rm p},3}^{(3)}\sim2\times10^{-7}\rm ~erg~cm^{-2}s^{-1}.\label{eicflux4}
\end{eqnarray}
Summarizing the above two cases, the EIC process of the
reverse-shocked electrons produces a TeV and a GeV emission
component, both of which are much weaker than the observed
synchrotron MeV emission with a flux of $\sim2.3\times10^{-5}\rm
~erg~cm^{-2}s^{-1}$.

 (iii) For $i=2$ and $j=2$, we have
\begin{eqnarray}
h\nu_{L,2}^{(2)}=2{{\gamma}^{\dag2}_{e,c}}h\nu_{m,2}=0.5~{\rm
keV}~x_0^{-1/2}Y_{-0.7}^{-2}y_{3,0}^{-1}\epsilon_{B,-1}^{-1}\hat{\epsilon}_{e,-1}^{3},
\end{eqnarray}
\begin{eqnarray}
h\nu_{H,2}^{(2)}=2{{\gamma}^{2}_{e,m,2}}h\nu_{c,2}=10~{\rm
MeV}~x_0^{35/32}y_{2,0}^{-2}y_{3,0}^{35/16}\epsilon_{B,-1}^{3/16}\hat{\epsilon}_{e,-1}^{45/16},
\end{eqnarray}
and the peak flux at $\nu_{H,2}^{(2)}$
\begin{eqnarray}
\left[\nu F_{\nu}\right]_{{\rm p},2}^{(2)}\sim7\times10^{-7}\rm ~erg~cm^{-2}s^{-1}.\label{eicflux1}
\end{eqnarray}
Due to the stronger synchrotron MeV emission, this relatively weaker
MeV emission component is likely to be covered up.

(iv) For $i=2$ and $j=3$, we have
\begin{eqnarray}
h\nu_{L,2}^{(3)}=2{{\gamma}^{\dag2}_{e,c}}h\nu_{m,2}=0.3~{\rm
GeV}~x_0^{5/16}Y_{-0.7}^{-2}y_{3,0}^{-11/8}\epsilon_{B,-1}^{-11/8}\hat{\epsilon}_{e,-1}^{11/8},
\end{eqnarray}
\begin{eqnarray}
h\nu_{H,2}^{(3)}=2{{\gamma}^{2}_{e,m,2}}h\nu_{m,3}=0.9~{\rm GeV}~\hat{\epsilon}_{e,-1}^{2}.
\end{eqnarray}
Distinct from the emission components discussed in the previous three cases, the flux at $\nu_{H,2}^{(3)}$,
\begin{eqnarray}
\left[\nu F_{\nu}\right]_{{\rm p},2}^{(3)}\sim2\times10^{-5}\rm ~erg~cm^{-2}s^{-1},\label{eicflux2}
\end{eqnarray}
indicates a strong GeV emission that is as strong as the synchrotron
MeV emission. This means that the energy of the forward-shocked
electrons would be mainly released by upscattering the synchrotron
MeV $\gamma$-ray photons from the reverse shock, while the
reverse-shocked electrons lose a great part of their energy via
synchrotron cooling directly. Due to the strong emission within the
energy regime from sun-GeV to GeV, electron-positron pairs might be
produced by collisions between the sub-GeV and GeV photons.
According to the results obtained above, we can give approximately
an upper limit luminosity of $L_{\rm lim}\sim10^{53}~\rm erg~s^{-1}$
for the sub-GeV ($\varepsilon_{\gamma}\sim0.1\rm GeV$) emission.
Then, the optical depth due to pair production interactions can be
roughly estimated as
\begin{eqnarray}
\tau_{\gamma\gamma}\lesssim{3\over8}\sigma_T{L_{\rm lim}\over4\pi R^2
c\gamma\varepsilon_{\gamma}}{R\over\gamma}\sim0.3~L_{\rm lim,53}\gamma_{2.6}^{-2}R_{16}^{-1},
\end{eqnarray}
which indicates that the pair production effect is not significant.
For simplicity, the further contribution from the secondary
electrons is ignored here.

To summarize, the contributions by the SSC and EIC emission to the
observed optical and MeV $\gamma$-ray emissions are insignificant
and the two synchrotron components are dominant in the observed
bands for GRB 080319B. In contrast, some higher-energy emission
components would be produced by the EIC process. Although most of
these components are weak, the flux of the strong GeV emission can
reach as high as $\sim10^{-5}\rm erg~cm^{-2}~s^{-1}$, which is
comparable to the observed MeV one.

\section{Summary and conclusions}
In the popular internal shock model for the prompt emission of GRBs,
paired forward and reverse shocks are produced by collisions between
some relativistic shells with different Lorentz factors. In this
paper, we have considered this model in an unusual situation where a
bimodal distribution of the shell Lorentz factors exists. As a
result, the Lorentz factors of the forward and reverse shocks are
quite different (i.e., the forward shock is Newtonian and the
reverse shock is relativistic) and the resulting two-component
synchrotron emission is expected to provide a new scenario for some
seldom GRBs that have two spectral peaks in the prompt emission.

As an example, we compare our scenario with the recently-observed
naked-eye GRB 080319B and constrain the model parameters by fitting
the observations. We find that, on one hand, the optical and MeV
$\gamma$-ray fluxes of this unique GRB could be explained in our
two-component synchrotron emission scenario, if some unknown
physical processes of the central engine (e.g., the picture
described in \S2.1) can give rise to a fireball where the Lorentz
factors of $\sim400$ and $\sim10^5$ appears alternately in a
variability timescale of few seconds. On the other hand, although
the internal shock-produced emission can roughly account for the
fluctuating structure of the light curves and the mild temporal
correlation between the optical and $\gamma$-ray emissions, it is
still difficult to explain clearly why the observed optical emission
varies relatively slower than the $\gamma$-ray emission. We
speculate that this difference between the light curve variabilities
may be due to a complicated distribution of Lorentz factors in the
fireball, an inhomogeneous structure of each fireball shell,
different shock-crossing times, and other more realistic properties
of the system. So, a more detailed simulation is required to improve
our present model.

Finally, for high energy emission, the synchrotron plus SSC scenario
suggested by Kumar \& Panaitescu (2008) and Racusin et al. (2008b)
predicts significant GeV $\gamma$-ray emission by considering the
second order IC-scattering, the flux of which is about 10 times
higher than the observed MeV one. In contrast, our model predicts a
relatively weaker GeV component, whose flux is lower than or at most
comparable to that of the synchrotron MeV emission. Therefore,
future observations for high energy counterparts of GRBs by the
\textit{Fermi} Space Telescope are expected to be able to
discriminate these two models.

\acknowledgements We would like to thank the referees for their
comments that have allowed us to improve this paper. This work is
supported by the National Natural Science Foundation of China
(grants no. 10221001, 10640420144, 10403002 and 10873009) and the
National Basic Research Program of China (973 program) No.
2007CB815404 and 2009CB824800. YWY is also supported by the
Scientific Innovation Foundation of Huazhong Normal University and
the Visiting PhD Candidate Foundation of Nanjing University.

\end{document}